\documentclass[a4paper,12pt]{article}

\usepackage{amsmath}
\usepackage{amssymb}
\usepackage{natbib}
\usepackage{graphicx}
\usepackage{subfigure}

\usepackage{amsfonts, amsthm, amsmath, amssymb}
\usepackage{natbib}
\usepackage{threeparttable}
\usepackage{fullpage}
\usepackage{setspace}
\usepackage[displaymath,mathlines]{lineno}
    \newcommand*\patchAmsMathEnvironmentForLineno[1]{%
      \expandafter\let\csname old#1\expandafter\endcsname\csname #1\endcsname
      \expandafter\let\csname oldend#1\expandafter\endcsname\csname end#1\endcsname
      \renewenvironment{#1}%
         {\linenomath\csname old#1\endcsname}%
         {\csname oldend#1\endcsname\endlinenomath}}%
    \newcommand*\patchBothAmsMathEnvironmentsForLineno[1]{%
      \patchAmsMathEnvironmentForLineno{#1}%
      \patchAmsMathEnvironmentForLineno{#1*}}%
    \AtBeginDocument{%
    \patchBothAmsMathEnvironmentsForLineno{equation}%
    \patchBothAmsMathEnvironmentsForLineno{align}%
    \patchBothAmsMathEnvironmentsForLineno{flalign}%
    \patchBothAmsMathEnvironmentsForLineno{alignat}%
    \patchBothAmsMathEnvironmentsForLineno{gather}%
    \patchBothAmsMathEnvironmentsForLineno{multline}%
    }

\newcommand{\R}{{\mathbb{R}}}

\title{Bayesian inference for latent factor GARCH models}
\author{Michael K. Pitt\\{\small Economics Department}\\{\small University of Warwick}\\{\small m.pitt@warwick.ac.uk}
\and Jamie Hall\\{\small School of Economics}\\{\small University of New South Wales}\\{\small jamie1212@gmail.com} \and Robert Kohn\\{\small School of Economics}\\{\small University of New South Wales}\\{\small r.kohn@unsw.edu.au}}
\date{\today}

\begin{document}
\maketitle

\begin{abstract}
Latent factor GARCH models are difficult to estimate using Bayesian methods because standard Markov chain Monte Carlo  samplers produce slowly mixing and inefficient
draws from the posterior distributions of the model parameters. This paper describes how to apply the particle Gibbs algorithm to estimate factor GARCH models efficiently. The method has two advantages over previous approaches. First, it
generalises in a straightfoward way to models with multiple factors and to various members of the GARCH family. Second,
it scales up well as the dimension of the observation vector increases.
{\newline}
\noindent Keywords: Particle Gibbs; Reversible jump
{\newline}JEL codes: C11, C38.
\end{abstract}

\section{Introduction}

This paper discusses latent factor generalised autoregressive conditionally heteroscedastic (GARCH) models. Factor GARCH models play two roles. The first is that they are a convenient type of multivariate GARCH model \citep{bauwens_multivariate_2006}, in which the factor structure provides a direct and parsimonious way to model the effects of time-varying volatility in a multivariate setting. Second, they are a natural extension of the latent factor model approach to time series modelling. For instance, factor models have a good track record in macroeconomic time series analysis \citep{stock_forecasting_2002}, and the addition of GARCH errors could improve a model's fit when the observations are conditionally heteroscedastic.

The use of factor GARCH models in a Bayesian context has been hampered by their computational difficulty. While it is possible to use single-move MCMC methods for this class of models, the resulting draws only explore the posterior distribution of the parameter vector  slowly and inefficiently \citep{fiorentini_likelihood-based_2004}.
These methods are also difficult to implement in models with more than one latent factor.

The main contribution of our article is to demonstrate that the particle Gibbs sampler \citep{andrieu_particle_2010} is particularly well suited to nonlinear latent factor models with a GARCH structure because we can use a fully adapted particle filter \citep{pitt_filtering_1999}  to produce rapidly mixing draws from the parameter vector. 
Our article shows that the method can handle multiple factors and can be
easily be generalised to apply to other members of the GARCH family. The statistical structure of factor GARCH models means that
the method scales well with the dimension of the observation vector because the variability in the estimated likelihood due to generating the latent factors decreases
as the dimension of the observation vector increases. Our article also shows that in certain cases we can make the inference invariant to the order of the elements in the vector of the dependent variable, while in the general case we take care to ensure that the empirical results are invariant to the order.
Although not shown in this article, it is clear that our method will accommodate regime changes and structural breaks in a straightforward way.

The paper is organized as follows. Section~\ref{section:inference} outlines the model and the sampling scheme. Section~\ref{s: simulated examples} demonstrates the performance of the sampling scheme using a simulated example. Section~\ref{s: empirical example} applies  the methodology to US stock returns.

\section{Inference} \label{section:inference}
We consider a factor model given by
\begin{align}
y_t & = \beta f_t + \epsilon_t ,
\label{eq: main model}
\end{align}
where $y_t$ is an $N \times 1$ vector of observations, $f_t$ is a $K \times 1$ vector of latent factors (where $K \ll N$), and $\epsilon_t$ is an $N \times 1$ vector of idiosyncratic errors. The distributions of $f_t$ and $\epsilon_t$ can be chosen in many ways. Our focus is on versions of (\ref{eq: main model}) in which the latent factors have volatilities given by GARCH processes, so that
\begin{align}
f_{j,t} &\sim N(0, \lambda_{j,t}^F)  \\
\label{eq: lambda update} \lambda_{j,t+1}^F &= \gamma_j + \alpha_j \lambda_{j,t}^F + \theta_j f_{j,t}^2 ,
\end{align}
for $j = 1, \dots, K$. In other words, $\lambda_{j,t}^F$ represents the conditional variance of the $j^{th}$ factor. We represent the diagonal variance-covariance matrix as $\Lambda^F_t$, where the $j^{th}$ entry on the diagonal is $\lambda_{j,t}^F$.

The methods described here can be applied to models with many different distributions over the idiosyncratic errors $\epsilon_t$. In what follows, we will assume that 
$\epsilon_t \sim N(0,\Lambda_t^E)$, with $\Lambda_t^E$ a diagonal matrix. Within this framework, we will consider two broad cases. The first is when  $\Lambda_t^E$ is
is the same unknown parameter $\Lambda^E$ at all time periods; in the second case, the idiosyncratic variances can follow GARCH processes similarly
 to the variances of the latent factors:
\begin{align}
\Lambda_t^E &= \mathrm{diag}\left(\lambda_{1,t}^E, \dots, \lambda_{N,t}^E\right) \\
\label{eq: lambda E update}
\lambda{i,t}^E &= \delta_i + \rho_i \lambda_{i,t-1}^E + \phi_i \epsilon_{i,t-1}^2
\end{align}
for $i = 1, \dots, K$. Conditional on information up to period $(t-1)$, using either assumption regarding the behaviour of $\Lambda_t^E$, the vector of latent factors is distributed as $f_t \sim N(0,\Lambda^F_t)$, and the observation vector $y_t \sim N(0, \beta \Lambda^F_t \Lambda^{F\prime}_t \beta^\prime + \Lambda_t^E)$. We show below 
that this conditionally Gaussian structure makes sequential Monte Carlo inference particularly efficient because a fully adapted particle filter can be applied.

The free parameters of the model are the elements of $\beta$, the factor GARCH parameters $(\gamma_j, \alpha_j,\theta_j)$ for $j = 1, \dots, K$, and either the fixed values of $\lambda_i^E$ or the GARCH pararameters $(\delta_i, \rho_i, \phi_i)$ for the idiosyncratic errors indexed by $i = 1, \dots, N$. We can carry out inference on these parameters, and generate forecasts of future observations, using the Gibbs sampler. The following sections briefly outline how an efficient Gibbs sampler can be applied to the model in (\ref{eq: main model}).

Note that the estimation method described below can be generalised straightforwardly to any variant of the model which is conditionally linear and Gaussian (that is, conditional on the latent state at time $t-1$). For instance, we use a GARCH-in-mean (GARCH-M) model for the observation vector in the empirical application described in detail below. In general, the methods described here can be applied to many members of the GARCH family. It is also straightforward to include exogenous observed variables on the right-hand side of (\ref{eq: main model}), though we omit this option throughout the paper for clarity.

\subsection{The latent factors}

Conditional on the parameters of the model, we can draw of the latent factors $f_{1:T}$ using a fully adapted variant of particle Gibbs \citep{andrieu_particle_2010}. To improve the efficiency of the Gibbs draws, we implement ancestor sampling \citep{lindsten_ancestor_2012}, as described below. In addition to conditioning on the parameters, the particle Gibbs algorithm also uses a draw of $f_{1:T}$ from a previous iteration. The sampler is initialised by setting this previous draw to an
arbitrary value. We begin the particle Gibbs algorithm with $M$ copies of the factor variance in the first period, $\Lambda^F_1$, initialised to its unconditional value,
\begin{align*}
\mathrm{diag}\left(\frac{\gamma_j}{1-\alpha_j-\theta_j}\right).
\end{align*}
Additionally, the first particle takes the value of $f_1$ from the previous draw $f_{1:T}$ that we
 condition on, and the remaining $(M-1)$ particles get a draw of $f_1 \sim N(0, \Lambda^F_1)$. Conditional on the first observation $y_1$, we resample the particles in proportion to their likelihood
 \begin{align*}
 p(y_1 | f_1) = \propto \exp \left( -\sum\frac{y_{1i} - \beta_i f_1}{2 \lambda^E_{i,1}} \right).
 \end{align*}
We then choose a particle index $b_1$ using an ancestor sampling step, described below.

For the following periods $t \in \{2, \dots, T\}$, we carry out the following steps:
\begin{enumerate}
\item Calculate the one-step prediction weights $\omega_{t|t-1}^{(k)} = p(y_t | f_{t-1}, \Lambda^F_{t})$ for $k \in \{1, \dots, M\}$,
\[ \omega_{t|t-1}^{(k)} = \frac{1}{(2\pi)^{\frac{n}{2}}} \left| W^{(k)} \right|^{-\frac{1}{2}} \exp \left( -\frac{1}{2} y_t^{\prime} (W^{(k)})^{-1} y_t \right)  \]
where
\begin{align*}
  (W^{(k)})^{-1} &= \left( (\Lambda^E_t)^{-1} - (\Lambda^E_t)^{-1} \beta (H^{(k)})^{-1} \beta^\prime (\Lambda^E_t)^{-1} \right) \\
  (H^{(k)})^{-1} &= \beta^\prime (\Lambda^E_t)^{-1} \beta + (\Lambda_{t}^{F})^{-1}
  \end{align*}
and the values of $\Lambda^E_t$ and $\Lambda^F_t$ are conditional on the value of the particle indexed by $k$.
\item Resample the particles $1, \dots, M$ with probability $\omega_{t|t-1}^{(k)} / \sum_j \omega_{t|t-1}^{(j)}$, but keep the particle indexed by $b_t$ unchanged.
\item Draw a value of $f \sim p(f_t | y_t, f_{t-1}, \Lambda^F_{t})$ for each particle $k \in \{1, \dots, M\}$. This density can be calculated as $f^{(k)} \sim N(\mu_t^{(k)}, H^{(k)})$, where
 \[ \mu_t^{(k)} =  (H^{(k)})\beta^\prime \Lambda^E_t y_t . \]
 Keep the particle indexed by $b_t$ unchanged.
\item Update the values of $\Lambda_{t+1}^{F}$ and $\Lambda_{t+1}^{E}$ for each particle, using equations (\ref{eq: lambda update}) and (\ref{eq: lambda E update}).
\item Carry out an ancestor sampling step by calculating the backward weights $w_{t|T}^{(k)}$ given by
\begin{equation} w_{t|T}^{(k)} = \omega_{t|t-1}^{(k)} \prod_{s={t+1}}^T p(y_s | f_{s}, \Lambda^F_{s}) p(f_s | \Lambda^F_{s}),
\label{eq: ancestor weights}
\end{equation}
where we condition on the particular draw $f_t^{(k)}$ and associated $\Lambda^F_{t+1}$ for the first period, then on the given path $\widehat{f}_{(t+1):T}$ thereafter. Following \cite{lindsten_ancestor_2012}, we truncate the product in (\ref{eq: ancestor weights}) after a fixed number periods. (In the examples below, we truncate after five periods, since this provided satisfactory approximations to the exact values.) As a result, the computing time increases linearly with $T$, rather than quadratically.
\item Choose a particle index $b_t \propto w_{t|T}$, and set $\widehat{f}_t = f_t^{(b_t)}$.
\end{enumerate}

\subsection{The factor loadings}  \label{ss: factor loadings}

Conditional on a draw of $f_{1:T}$, we can take a draw of the factor loadings $\beta$. Broadly, there are two approaches that can both be used at this point. The first approach, widely used in Bayesian inference, involves imposing restrictions on the structure of $\beta$ to guarantee identification. A well-known aspect of factor models such as (\ref{eq: main model}) is that we cannot directly identify a single set of values for the latent factor series $f_{1:T}$, but only an equivalence class under the action of orthogonal rotations. In other words, if $Q$ is any $K\times K$ orthogonal matrix, then a vector of factors $f_{t}$ with loadings $\beta$ will have the same likelihood as another vector $\widetilde{f}_t = Qf_t$ with loadings $\widetilde{\beta}=\beta Q^\prime$. In order to pin down particular values for the latent factors, we therefore need extra identifying assumptions. Two common choices in the Bayesian econometric literature are, first, to let $\beta$ be a triangular matrix and assume that $f$ has unit variance \citep{geweke_measuring_1996}; or, second, to have an unrestricted variance for $f$ and assume that $\beta$ is triangular with ones on the diagonal \citep{aguilar_bayesian_2000}. Using one of these assumptions, and given a draw of $f_{1:T}$, the posterior distribution of $\beta$ is conditionally normal. Taking a conditional draw of $\beta$ effectively means carrying out a linear regression. Thus it is straightforward to apply one of these identification schemes in the Gibbs sampler, and we use one of them for our GARCH-M example below.

This type of identification scheme is simple, generally applicable, and widely used. Its main drawback is that the resulting inference can depend on the order in which the components of $y_t$ happen to be arranged \citep{chan_efficient_2012}. The reason for this is that the triangular identification schemes introduce a discontinuity into the mapping from the reduced-form values $\beta f_t$ to the factor loadings $\beta$. One can understand this intuitively by considering arrangements in which the $i^{th}$ observation is in fact uncorrelated with the $i^{th}$ factor. The triangular identification schemes effectively place a prior weight of zero on those cases. Thus their assessment of the model's fit to the data can be severely hampered. Instead, it is possible to use a sampling method that is invariant to reorderings of the observation vector \citep{chan_efficient_2012}, which we summarise here. The disadvantage of this method is that at present it may not be
as widely applicable; its assumptions are invalidated if the latent factors have a GARCH-M structure instead of GARCH, for instance. We must also assume in this case that the idiosyncratic errors have constant variance, $\Lambda^E_t = \Lambda^E$. These restrictions may or may not be important, depending on the application in question. We use this method for our simulated examples in Section~\ref{s: simulated examples} where we assume constant error variances, as well as using it to provide a robustness check in the more general model of
Section~\ref{s: empirical example}

Briefly, the invariant method for inference on $\beta$ first stacks
the row vectors $f_t^\prime$ into a matrix $F$. The matrix $F \beta^\prime$, which has rank $K$, can be decomposed as
\[ F \beta^\prime = U \Lambda, \]
where $U \in V_{K,T}$ and $\Lambda \in \R^{T \times n}$. Here $V_{K,T}$ is a Stiefel manifold. A Stiefel manifold is a space consisting of orthogonal $K$-frames in the ambient space $\R^T$ \citep{james_normal_1954,strachan_bayesian_2004}. The prior for $\Lambda$ is chosen to be matrix-normal, so that each element of $\Lambda$ is independently normally distributed with mean zero and variance $1/(c_\lambda)^2$. The matrix $U$ functions as the coordinates of the row space $sp(F)$ of the reduced-rank matrix $F \beta^\prime$, seen as an element of $G_{K,T}$, the Grassmannian $G_{K,T}$ being the space of all linear {$K$-dimensional} subspaces embedded in $R^T$ \citep{james_normal_1954,strachan_bayesian_2004}. The decomposition of $F \beta^\prime$ into $U \Lambda$ is not convenient enough to work with, because $U$ has a fixed orientation within the plane $sp(F)$ as the plane moves through $G_{K,T}$. However, suppose we act on it with an orthogonal matrix $C \in O(K)$ (writing $O(K)$ for the space of orthogonal matrices), so that
\[ U \Lambda = \left( U C \right) \left( C^\prime \Lambda \right) = U_a \Lambda_a \]
Then the matrix $\Lambda_a$ will have the same prior as $\Lambda$, by the rotational invariance of the normal distribution; and if $U$ and $C$ have uniform priors on $G_{K,T}$ and $O(K)$, then their product will have a uniform prior on $V_{K,T}$, given by $U^\prime dU = 1/(c_G c_O)$, where $c_G$ and $c_O$ are the normalising constants for $G_{K,T}$ and $O(K)$ \citep{james_normal_1954}.

Finally, the \cite{chan_efficient_2012} procedure introduces parameter expansion \citep{liu_parameter_1999} to turn the prior distribution into a convenient conjugate form. Let $A \sim W(I_r, T-N)$, and let $\kappa$ be its Cholesky decomposition, so that $A = \kappa \kappa^\prime$. Here, $W(A, \nu)$ is a Wishart distribution with
scale matrix $A$ and $\nu$ degrees of freedom.
Rewrite the reduced-form matrix as
\[ U_a \Lambda_a = (U_a \kappa) (\kappa^{-1} \Lambda_a) = F \beta^\prime . \]
The Jacobian for this transformation is
\[ (dA)(U_a^\prime dU_a ) (d\Lambda_a) = 2^K \left| F^\prime F \right|^{-(T-N-K-1)} (d\Lambda)(dF) \]. 
Based on the assumptions made so far, the prior is
\begin{align*} p(U_a, \Lambda_a, A) &= \left[ \frac{1}{c_G c_O} \left( \frac{c_\lambda}{2\pi} \right)^{NK/2} \exp \left( -\frac{c_\lambda}{2} \mathrm{tr} \Lambda_a \Lambda_a^\prime \right) (d\Lambda_a) \right. \\ & \hspace{1cm} \left. \frac{1}{c_W} \left| A \right|^{(T-N-K-1)/2} \exp\left( -\frac{1}{2} \mathrm{tr} A \right) (dA) \right], \end{align*}
with $c_W$ representing the normalising constant for the Wishart distribution. From this, it follows that
\[ p(\beta, F) \propto \left( \frac{c_\lambda}{2\pi} \right)^{NK/2} \exp \left( -\frac{c_\lambda}{2} \mathrm{tr} \beta F^\prime F \beta^\prime  \right) \exp \left( -\frac{1}{2} \mathrm{tr}F^\prime F \right) (d\beta)(dF) \] . 

Therefore, conditional on $F$, the prior on $\beta$ is Gaussian. Thus, the parameter expansion introduced by \cite{chan_efficient_2012} produces a conjugate prior for the regression of $\beta$ on $F$, meaning that the conditional posterior $p(\beta | F, \dots)$ is Gaussian. Note that, while the original analysis by \cite{chan_efficient_2012} uses homoscedastic latent factors, that assumption seems not to be required for this derivation to go through; in particular, it still applies when the rows of $F$ have time-varying volatility governed by a GARCH process.
Writing $\beta_i$ for the $i^{th}$ row of the loading matrix, and $y_{i}$ for the $T\times 1$ vector of observations on the $i^{th}$ data series, we have
\begin{align}
\nonumber
p(\beta_i | F, \Lambda^E, c_\lambda) \sim N\left( \rule{0pt}{10pt} \right. & \left[ (1 + c_\lambda \Lambda^E_{ii} ) F^\prime F \right]^{-1} F^\prime y_{i} \rule{3pt}{0pt},  \\ &  \rule{0pt}{3pt} \left. \Lambda^E_{ii} \left[ (1 + c_\lambda \Lambda^E_{ii} ) F^\prime F \right]^{-1}  \right)
\label{eq: invariant beta regression}
\end{align}
Although the \cite{chan_efficient_2012} method can be used to generate draws of $\beta$, which then provide conditional draws of $F$, it does not circumvent the original problem of separately identifying the rotation of $\beta F$. In other words, the draws of $\beta$ and $F$ generated by the Gibbs sampler are implicitly providing draws of the reduced-rank matrix $\beta F$. If an econometrician wishes to interpret the latent factors $F$ or the loadings $\beta$ separately, they must still impose some kind of identification scheme, such as the diagonal structures of \cite{geweke_measuring_1996} or \cite{aguilar_bayesian_2000}. The advantage of the \cite{chan_efficient_2012} approach is that the estimates of $\beta F$, and the resulting judgements about the accuracy of the model, do not depend on the order of the components of $y$. That would not be the case if we imposed the identification scheme \emph{during} the estimation.

\subsection{The idiosyncratic errors}

If the idiosyncratic error variances $\Lambda^E_{ii}$ are homoscedastic, then it is convenient to impose independent inverse Gamma prior distributions with mean $\mu_e$ and degrees of freedom $\nu_e$. Consequently, their posterior distributions are inverse Gamma with $(T+\nu_e)$ degrees of freedom and mean $\left(\sum_t \epsilon_{i,t}^2 + \nu_e/\mu_e\right)$.

If the idiosyncratic error variances are assumed to follow GARCH processes, then we can carry out inference on their parameters in the same manner as described in the next subsection.

\subsection{The GARCH parameters}

Having obtained draws of $f_{1:T}$ and $\beta$, we can use a Metropolis-within-Gibbs step to obtain draws of the GARCH parameters. To increase the efficiency of the Metropolis Hastings moves, we first reparameterise the set of GARCH parameters for factor number $j$ by
\begin{equation}  \psi_1 =  \alpha_j + \theta_j  \rule{20pt}{0pt} \psi_2 = \frac{\gamma_j}{1-\psi_1} \rule{20pt}{0pt} \psi_3 =  \frac{\alpha_j}{\psi_1}  \label{eq: garch repar1} \end{equation}
and then by
\begin{equation} \varphi_1 = \log(\psi_1) - \log(1-\psi_1) \rule{20pt}{0pt} \varphi_3 = \log(\psi_2) \rule{20pt}{0pt} \varphi_3 = \log(\psi_3) - \log(1 - \psi_3)  \label{eq: garch repar2}  \end{equation}
Writing $\varphi^c = (\varphi_1, \varphi_2, \varphi_3)$, we use these coordinates to propose new parameters via $\varphi^p \sim N(\varphi^c, \Sigma)$. The proposal covariance $\Sigma$ is initialised to a matrix with small positive numbers on the diagonal, and then updated using the adaptive Metropolis Hastings scheme of \cite{haario_adaptive_2001}. The new parameters $\varphi^p$ are accepted with probability
\[ p_j = 1 \wedge \left[ \frac{\pi(\widehat{f}_{j,1:T} | \varphi^p )}{\pi(\widehat{f}_{j,1:T} | \varphi^c )}\right]  \]
The conditional likelihood $\pi(\cdot)$ is given by
\[ \pi(\widehat{f} | \varphi) =  \prod_{t=1}^T \frac{1}{\sqrt{2 \pi \sigma^2_t}} \exp \left( -\frac{\widehat{f}_{j,t}^2}{2 \lambda_t} \right)  \]
where the variance follows
\[ \lambda_{t+1} = \gamma_j + \alpha_j \widehat{f}_t^2 + \theta_j \lambda_t \]. 
Here the proposed GARCH parameters are calculated from $\varphi^p$ by inverting (\ref{eq: garch repar1}) and (\ref{eq: garch repar2}).

When using the reordering-invariant specification for the factor loadings, the marginal variance of the latent factors becomes unidentified. In our implementation, we fix the marginal variance at 1, by assuming that $\gamma_j = (1-\alpha_j-\theta_k)$. This restriction is straightforward to impose in the Metropolis Hastings steps.

\subsection{The number of latent factors}  \label{ss: number of latents}

Inference on the number of factors $K$ can be carried out using the Reversible Jump method \citep{green_reversible_1995}. The Reversible Jump implementation described by \cite{lopes_bayesian_2004} for the homoscedastic factor case is  straightforward to extend 
 to models in which the latent factors and idiosyncratic variances follow GARCH processes. In this section, we summarise and explain this extension of the Reversible Jump method. We note in passing that models with homoscedastic latent factors can use the Savage-Dickey Density Ratio (SDDR) \citep{verdinelli_computing_1995}, as described by \cite{chan_efficient_2012}. However, after applying that alternative method to simulated examples, we concluded that it is not feasible for GARCH factor models, since the changing volatilities of the latent factors make the estimation of the SDDR inefficient and slow to converge.

In implementing the Reversible Jump method in the style of \cite{lopes_bayesian_2004}, we begin with separate preliminary estimation runs for all values of $K$ up to a chosen maximum $\overline{K}$. The draws of the factor loadings and the GARCH parameters from these preliminary runs are then used to generate proposal draws for the model-choice steps, as follows. Let $\overline{\beta}_k$ and $B_k$ denote the posterior mean and covariance of the factor loadings estimated from the preliminary estimation run with $k$ latent factors. The proposal density for $\beta$, conditional on using $k$ latent factors, is then $q_k(\beta) = \mathcal{N}(\overline{\beta}_k, b B_k)$, where $b > 0$ is a scale factor chosen to ensure that the tails of the proposal density are fat enough. Similarly, let the estimated posterior mean of the GARCH parameters, transformed as in equation (\ref{eq: garch repar2}), be denoted $\overline{\varphi}_k$, and the estimated posterior covariance be $\Phi_k$. The proposal density for the transformed GARCH parameters is then $q_k(\varphi) = \mathcal{N}(\overline{\varphi}_k, c \Phi_k)$, with $c > 0$ a fixed scale parameter. Note that using the transformed values $(\varphi_1,\varphi_2,\varphi_3)$, rather than the original parameters $(\gamma,\alpha,\theta)$, ensures that the proposed GARCH parameters are always positive and within the stable region. The proposal density for the idiosyncratic error variance parameters, which we will write as $q_k(\sigma^E)$, can either be an inverse gamma in the homoscedastic case, or a Gaussian on the GARCH parameters transformed as in equation (\ref{eq: garch repar2}).

Writing $\theta_k$ for the entire parameter vector used in a model with $k$ latent factors, the proposal density is
\begin{equation}
q_k(\theta_k) = q_k(\beta) q_k(\varphi) q_k(\sigma^E)
\end{equation}

At the end of each Gibbs sampler iteration, we can then generate a between-model move as follows. We sample a proposed value $k^\prime$ uniformly from $\{1,\dots, \overline{K}\}$. We then use the fully-adapted particle filter to obtain a simulated value of the proposed likelihood $p(y|k^\prime, \theta_{k^\prime})$ and the current likelihood $p(y | k, \theta_k)$. This allows us to compute the Reversible Jump acceptance probability
\begin{equation}
\alpha = \min \left[ 1, \frac{ p(y|k^\prime, \theta_{k^\prime}) p(\theta_{k^\prime}|k^\prime) p(k^\prime) q_k(\theta_k) }{ p(y | k, \theta_k) p(\theta_k|k) p(k) q_{k^\prime}(\theta_{k^\prime}) } \right] . 
\end{equation}
Here $p(\theta_k | k)$ is the prior density on the parameter vector $\theta_k$, and $p(k)$ is the prior mass on the number of latent factors. With probability $\alpha$, the proposed value $k^\prime$ is accepted, and we carry out the next Gibbs iteration using the proposed parameters $\theta_{k^\prime}$. In that case, the current draw of the latent factors $f_{1:T}$ can be padded with zeros if $k^\prime > k$, or truncated if $k^\prime < k$.

\section{Simulated examples} \label{s: simulated examples}

To evaluate the performance of the estimation method described above, we carried out a series of tests on simulated data. We simulated 200 periods of data using a model with two latent factors. For both factors' GARCH parameters, we chose $\alpha_j = 0.04$, $\theta_j=0.9$, and (as discussed above) $\gamma_j = (1-\alpha_j-\theta_j)$. Unless otherwise specified, we used $N=5$ observation components, $M=10$ particles, and idiosyncratic variances $\Lambda^E = \mathrm{diag}(0.02, \dots, 0.02)$. We carried out five replications of each experiment. That is, we simulated five different data sets using each combination of settings, then estimated the model on each one. For each replication, we obtained 20,000 parameter draws. In our estimation, we constrained the draws of $\gamma_j$ to be equal to $(1-\alpha_j-\theta_j)$ as discussed above. The diagonal values of $\Lambda^E$ (which are in fact the same) were estimated independently.

We evaluated the efficiency of the estimation method by calculating the integrated autocorrelation times (IACTs) for particular groups of parameters. Given a vector of draws $\theta_i$, the IACT is defined as
\begin{equation}
IACT(\theta) = 1 + 2 \sum_{\tau=1}^\infty \rho_\tau (\theta), 
\end{equation}
where $\rho_\tau(\theta)$ is the autocorrelation between $\theta_i$ and $\theta_{i+\tau}$. We calculated the IACTs using the overlapping batch means method of \cite{flegal_batch_2010}. The times can be intuitively interpreted as inflation factors relative to independent draws from a parameter's marginal posterior distribution. That is, a value of 20 would suggest that we require 20 draws from the algorithm to obtain the equivalent of one independent draw from the posterior.

\subsection{Results}

\begin{figure}
\centering
\mbox{\subfigure[Trace plot; black line shows the true value]{\includegraphics[width=5cm,trim=1cm 5cm 1cm 5cm]{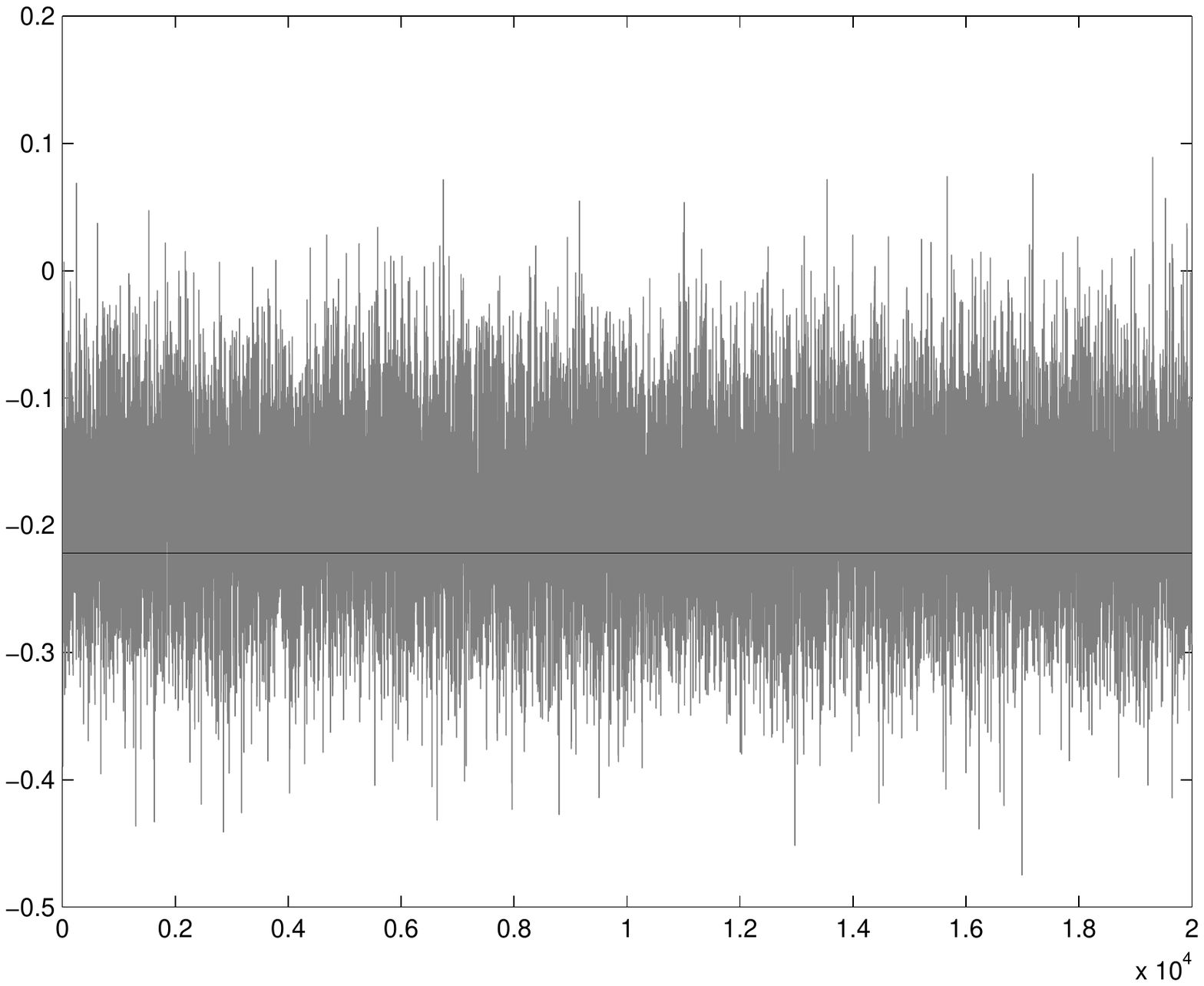}}\quad
\subfigure[Estimated autocorrelations]{\includegraphics[width=5cm,trim=1cm 5cm 1cm 5cm]{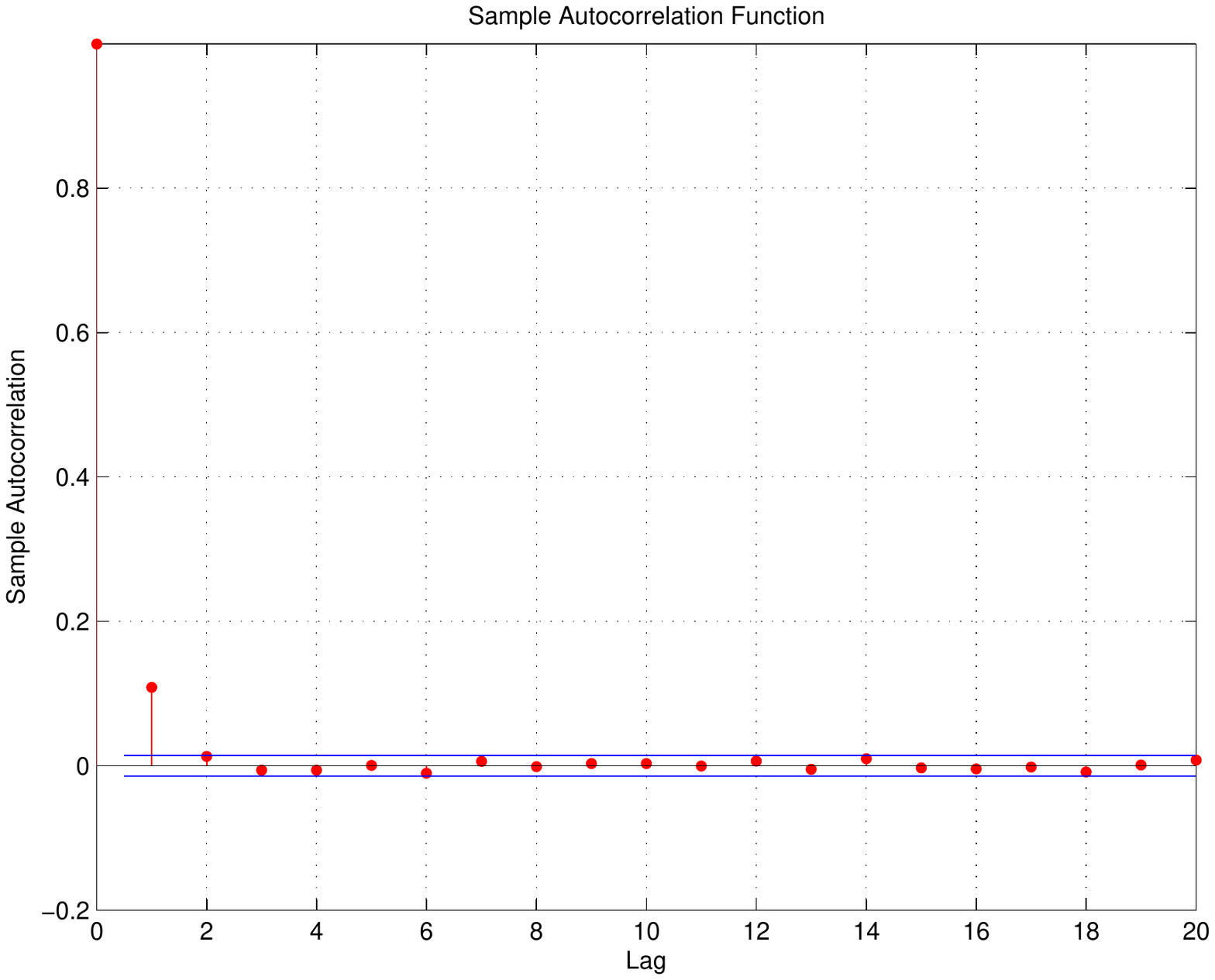} }}
\caption{Analysis of draws of  $(\beta F)_{1, 5}$from one run of the simulated example} \label{fig:trace and autocorr}
\end{figure}

The sampler produced parameter draws with a low degree of autocorrelation. The left panel of Figure~\ref{fig:trace and autocorr} shows
a typical trace plot of the draws of $\beta f$---in this case, corresponding to the first observational component in the fifth time period. It appears that the sampler is exploring the posterior distribution quite rapidly. This impression is supported by the estimated autocorrelations of those draws, presented in the right-hand panel of the figure.

We carried out the Reversible Jump procedure to estimate the number of latent factors. It
 selected two latent factors (the correct number) with a high probability.

\subsection{The number of particles}

We tried varying the number of particles $M$ to ascertain its effect on the efficiency of the estimation. Table~\ref{table: M results}
reports the median integrated autocorrelation times for different groups of the model parameters over the five replications.
Table~\ref{table: M max results} reports the maximum IACTs for each parameter group. The results suggest that the worst-case performance is fairly good, even for small numbers of particles.

\begin{table}
\begin{center}
\begin{tabular}{c|cccc}
$M$  \rule{0pt}{12pt}  &  $(\beta f)$ & $\Lambda^E$ & $\alpha$ & $\theta$  \\
\hline
5 & 1.73 & 6.4 & 37.8 & 55.0 \\
10 & 1.30 & 3.2 & 26.4 & 19.4 \\
20 & 1.17 & 2.9 & 29.1 & 17.4 \\
40 & 1.08 & 2.2 & 23.6 & 15.3 \\
\hline
\end{tabular}
\caption{Median integrated autocorrelation times for parameter groups, using different numbers of particles $M$}
\label{table: M results}
\end{center}
\end{table}
\begin{table}
\begin{center}
\begin{tabular}{c|cccc}
$M$  \rule{0pt}{12pt}  &  $(\beta f)$ & $\Lambda^E$ & $\alpha$ & $\theta$  \\
\hline
5 & 97.11 & 74.9 & 69.9 & 94.7 \\
10 & 79.70 & 52.2 & 63.4 & 95.9 \\
20 & 76.35 & 36.1 & 68.3 & 93.1 \\
40 & 4.26 & 8.0 & 42.5 & 30.8 \\
\hline
\end{tabular}
\caption{Highest integrated autocorrelation times for parameter groups, using different numbers of particles $M$}
\label{table: M max results}
\end{center}
\end{table}

The estimated median autocorrelation times are relatively low for all groups of parameters, but particularly for the reduced-form factor values $\beta f$ and the idiosyncratic noise variances $\Lambda^E$. The values for $\beta f$ in Table~\ref{table: M results} are consistent with the asymptotic theory developed by \cite{pitt_properties_2012} for particle Metropolis Hastings: the IACTs decrease in proportion to $\exp(1/M)$. However, the levels of the IACTs are considerably lower than might be expected from a Metropolis Hastings estimation run.

\begin{table}
\begin{center}
\begin{tabular}{c|cccc}
$M$  \rule{0pt}{12pt}  &  $(\beta f)$ & $\Lambda^E$ & $\alpha$ & $\theta$  \\
\hline
5 & 8.66 & 32.1 & 189.0 & 274.8 \\
10 & 13.03 & 32.1 & 263.6 & 193.7 \\
20 & 23.45 & 58.8 & 582.1 & 348.3 \\
40 & 43.21 & 86.7 & 944.6 & 613.7 \\
\hline
\end{tabular}
\caption{Median computing times, calculated as $M \times IACT$ }
\label{table: M-ct results}
\end{center}
\end{table}

Since the time required for running the particle Gibbs algorithm scales roughly in proportion to $M$, we can use the product of $M$ and the IACT as a measure of computing time. These values, reported in Table~\ref{table: M-ct results}, suggest that around 5 to 10 particles is optimal for this model.

\subsection{The number of observations}

We varied the number of observations $N$ from 5 up to 50.
Table~\ref{table: N results} summarises the results. The inference on the latent factors (our main object of inference) remains broadly similar, though the efficiency of inference about the idiosyncratic errors improves as the number of observations increases. The efficiency of the estimates of the GARCH parameters becomes somewhat poorer for medium-sized $N$, partly reflecting the difficulty of separately identifying the $\alpha$ and $\theta$ parameters. However, the efficiency then improves for $N = 100$ 
as the increase in $N$ helps to reduce the noise in the estimated likelihood because of the GARCH structure. 

\begin{table}
\begin{center}
\begin{tabular}{c|cccc}
$N$  \rule{0pt}{12pt}  &  $(\beta f)$ & $\Lambda^E$ & $\alpha$ & $\theta$  \\
\hline
5 & 1.29 & 4.8 & 44.0 & 13.6 \\
10 & 1.30 & 3.2 & 26.4 & 19.4 \\
25 & 1.29 & 2.1 & 65.6 & 69.8 \\
50 & 1.32 & 2.4 & 60.8 & 61.1 \\
100 & 1.53 & 5.2 & 25.0 & 18.2 \\
\hline
\end{tabular}
\caption{Median integrated autocorrelation times for different parameters with various sizes of the observation vector $y_t$}
\label{table: N results}
\end{center}
\end{table}

\subsection{The idiosyncratic noise variance}

We also varied $\Lambda^E$, the variance of the idiosyncratic noise vector $\epsilon$. That is, the values on the diagonal of the $\Lambda^E$ matrix were set to the same value on each repetition, but we made that common value higher and lower, using the values listed in Table~\ref{table: noise results}.

\begin{table}
\begin{center}
\begin{tabular}{c|cccc}
$\Gamma$  \rule{0pt}{12pt}  &  $(\beta f)$ & $\Lambda^E$ & $\alpha$ & $\theta$  \\
\hline
0.10 & 6.15 & 44.7 & 27.6 & 39.1 \\
0.02 & 1.30 & 3.2 & 26.4 & 19.4 \\
0.01 & 1.28 & 3.3 & 40.9 & 31.5 \\
\hline
\end{tabular}
\caption{Median integrated autocorrelation times for different parameters with various idiosyncratic noise variances $\Lambda^E$}
\label{table: noise results}
\end{center}
\end{table}

Unsurprisingly, the performance of the sampler degrades somewhat as $\Lambda^E$ increases, meaning that the observations become less informative about the latent factors.

\section{Empirical application to US stock returns} \label{s: empirical example}

We applied the estimation method described above to a sample of monthly US stock returns. Our data, kindly provided by Kenneth French, consisted of the monthly returns for 17 value-weighted industry portfolios. We used a sample running from January 1980 to December 2012, a total of 396 observations.

In this case, we used a GARCH-M model for the volatility of the latent factors. This is equivalent to positing a leverage effect---that is, an interaction between the factors' volatilities and their conditional means. The model is summarised by
\begin{equation}
y_t = \beta \widetilde{f}_t + \epsilon_t ,
\label{eq: garch-m}
\end{equation}
where now the latent factors $\widetilde{f}_t$ have conditional means depending on their volatility
\begin{align}
\widetilde{f}_{j,t} &\sim N( \tau_j \lambda_{j,t}^F, \lambda_{j,t}^F) . 
\label{eq: garch-m lambda update}
\end{align}
Inference on this model can be carried out through a straightforward generalisation of (\ref{eq: main model}), since it maintains the conditionally linear and Gaussian structure of the basic factor GARCH model. The conditional variances $\Lambda_t^F$ and $\Lambda^E_t$ are assumed to have GARCH structures, as described in section \ref{section:inference}.

This model invalidates the assumptions of the invariant factor loading estimation method, so we chose instead to identify the factor loadings by assuming that the loading matrix $\beta$ has ones on the main diagonal and zeros above the main diagonal. We imposed independent $N(0,1)$ priors on the rest of the elements of $\beta$. For the leverage parameters, we assumed independent priors given by $\tau_j \sim N(0,1)$, and each of the GARCH parameters $\gamma_i$, $\alpha_i$, $\theta_i$, $\delta_i$, $\rho_i$ and $\phi_i$ were given independent $U[0,1]$ priors.

Given the difficulties identified by \cite{chan_efficient_2012}, the ordering of the components of the observation vector
 should be chosen with some care. Inference becomes unstable if the $i^{th}$ observation component is not, in fact, correlated with the $i^{th}$ latent factor. In the case of the dataset used here, we have no prior information about industry groups that would make one ordering seem more reasonable than another. We therefore chose an ordering in two stages, starting by roughly ranking the series in order of the explanatory power, and then confirming that the resulting order was not obviously inconsistent with an invariant specification.

For the first stage, we carried out a linear regression of each of the 17 industry return series on every other series, selecting the series that provided the highest single $R^2$ as our first observation component. We chose the second and subsequent series by recursively adding them to the set of independent variables in the linear regressions, each time choosing the series that provided the highest single $R^2$, and stopping after choosing the first eight components of the observation vector.

Having established this ordering, we checked it against the results of a factor GARCH model that we estimated with reordering-invariant loadings, as described above. (Note that this model assumes homoscedastic idiosyncratic errors and no GARCH-in-mean effect, so it is only an approximation of our final model specification; this check is indicative rather than conclusive.) Consider a single draw $\widetilde{\beta}$ of the factor loadings from the invariant specification of the model. We can relate it to the specification with ones on the main diagonal by using a QR decomposition of $\widetilde{\beta}$ to write
\begin{equation}
\widetilde{\beta} = \left[ \begin{array}{c} L \\ Y \end{array} \right] Q = \left[ \begin{array}{c} E \\ Z \end{array} \right] D Q = \beta D Q . 
\end{equation}
where $Q$ is orthogonal, $L$ is lower triangular, $E$ is lower triangular with ones on the diagonal, and $D$ is diagonal and positive. So, if the corresponding draw of the latent factors is $\widetilde{f}$, then the estimate of $f$ produced by the specification with ones on the diagonal can be calculated as $Q^{-1}D^{-1}\widetilde{f}$. This will be numerically unstable if the numbers on the diagonal of $D$ are too small. We therefore took 1000 draws using the invariant algorithm and the proposed ordering of variables, with the number of latent factors $K$ ranging from 1 to 8. The smallest element of $D$ was found to be $6 \times 10^{-5}$, which occurred in the eight factor case. While far from ideal, this is well within floating-point precision. This suggests that our variable ordering is adequate.

We estimated this model on the monthly equity return data using the Reversible Jump method described above to choose the number of latent factors. In the preliminary estimation runs, we estimated models with between one and eight latent factors.

\subsection{Results}

The Reversible Jump method placed a high probability on the version of the model with seven latent factors. In Table~\ref{table: us17 SNRs}, we report the estimated signal to noise ratios for each of the 17 observation components. The table reports the sample variance of each component (with the monthly returns measured in percentage points), and its estimated idiosyncratic variance. The latter was calculated as the mean of the unconditional noise variance $\Lambda^E_i$, given by ${\delta_i/(1-\rho_i-\phi_i)}$. The final column reports the signal to noise ratio, calculated as the difference between the sample variance and the idiosyncratic variance, divided by the idiosyncratic variance. Most components are estimated to have fairly high SNRs. This means that the fully adapted filter for the GARCH components will be very efficient in this case.

\begin{table}
\begin{center}
\begin{tabular}{l|ccc}
Industry sector \rule{0pt}{12pt}  &  \begin{tabular}[c]{@{}c@{}}Sample\\variance\\({\%})\end{tabular} & \begin{tabular}[c]{@{}c@{}}Idiosyncratic\\variance\\({\%})\end{tabular} & SNR    \\
\hline
Food \rule{0pt}{14pt} & 18.5 & 1.2 & 14.6 \\
Mining and Minerals & 65.9 & 29.1 & 1.3 \\
Oil and Petroleum Products & 33.1 & 4.2 & 6.8 \\
Textiles, Apparel {\&} Footwear & 39.1 & 15.0 & 1.6 \\
Consumer Durables & 33.0 & 7.7 & 3.3 \\
Chemicals & 34.9 & 7.2 & 3.8 \\
Drugs, Soap, Perfumes, Tobacco & 20.4 & 14.4 & 0.4 \\
Construction {\&} Construction Materials & 37.8 & 11.5 & 2.3 \\
Steel Works Etc & 66.6 & 49.2 & 0.4 \\
Fabricated Products & 31.4 & 6.8 & 3.6 \\
Machinery {\&} Business Equipment & 50.4 & 24.5 & 1.1 \\
Automobiles & 46.2 & 6.5 & 6.1 \\
Transportation & 30.6 & 7.3 & 3.2 \\
Utilities & 15.8 & 3.7 & 3.3 \\
Retail Stores & 28.0 & 7.9 & 2.5 \\
Banks {\&} Insurance Companies & 31.3 & 2.6 & 10.8 \\
Other  & 26.1 & 6.9 & 2.8 \\
\hline
\end{tabular}
\caption{Estimated variances for US industry sectors.}
\label{table: us17 SNRs}
\end{center}
\end{table}

We find no evidence for leverage effects in this dataset, with the posterior credible intervals of each $\tau_j$ including zero in all model variants. This stands in contrast to the results on UK stock return data analysed in \cite{fiorentini_likelihood-based_2004}, which used a single latent factor with homoscedastic idiosyncratic errors. It may be that putative leverage effects can appear as artefacts of time-varying idiosyncratic volatility.

\begin{figure}
\centering
\mbox{\subfigure[Trace plot]{\includegraphics[width=5cm,trim=1cm 5cm 1cm 5cm]{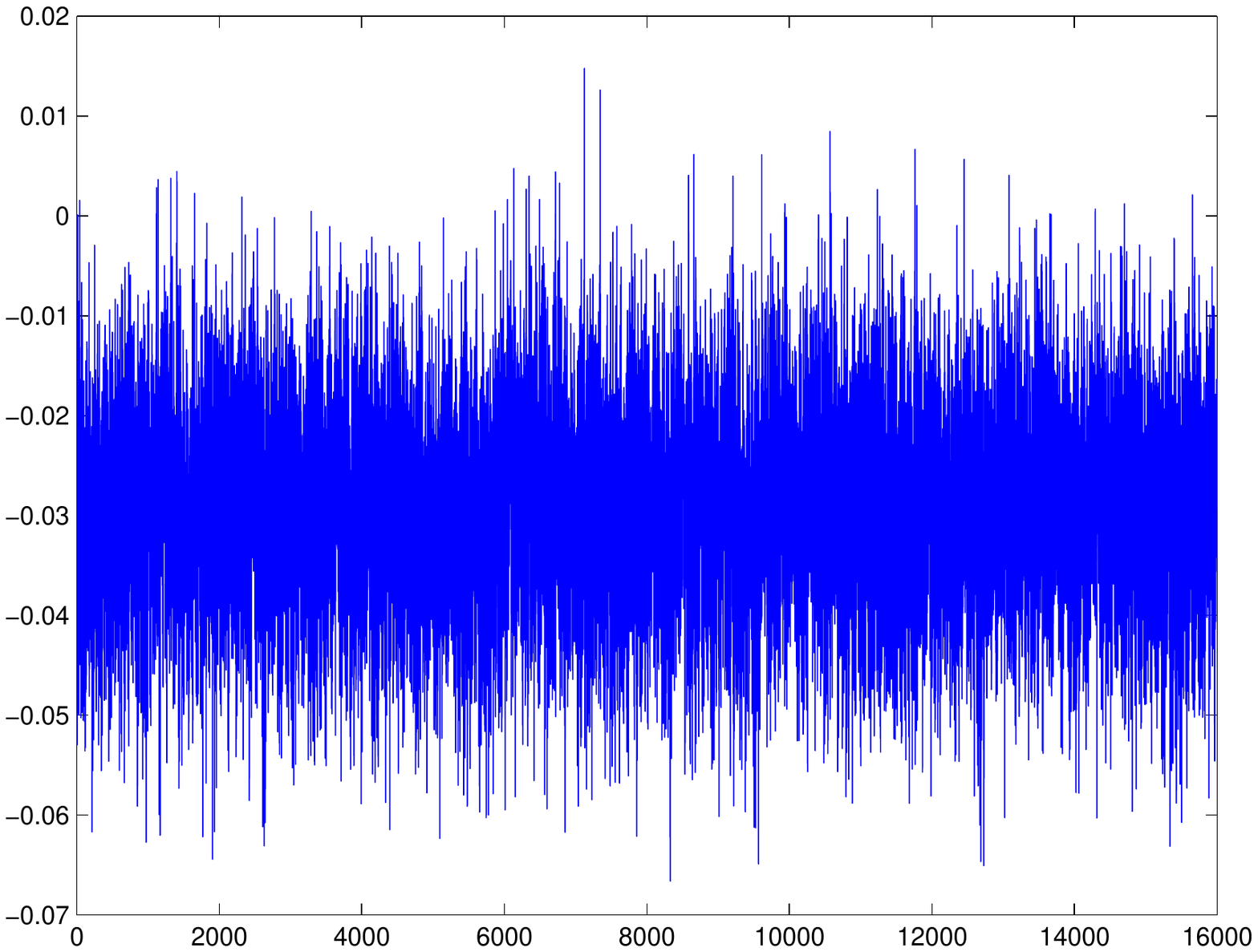}}\quad
\subfigure[Estimated autocorrelations]{\includegraphics[width=5cm,trim=1cm 5cm 1cm 5cm]{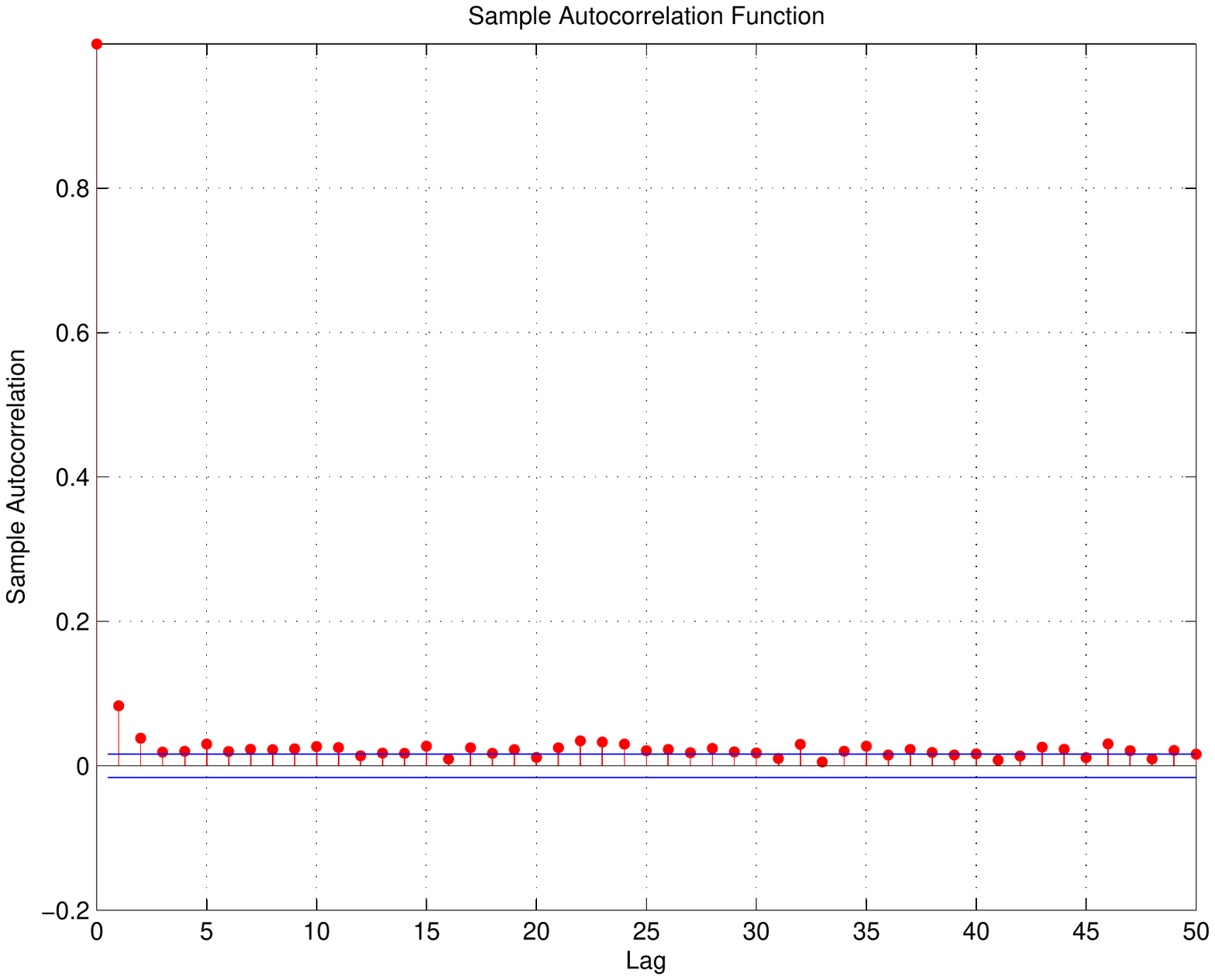} }}
\caption{Analysis of draws of $f_{150,1}$ from the estimated GARCH-M model with four latent factors} \label{fig:us17 f}
\end{figure}

\begin{figure}
\centering
\mbox{\subfigure[Trace plot]{\includegraphics[width=5cm,trim=1cm 5cm 1cm 5cm]{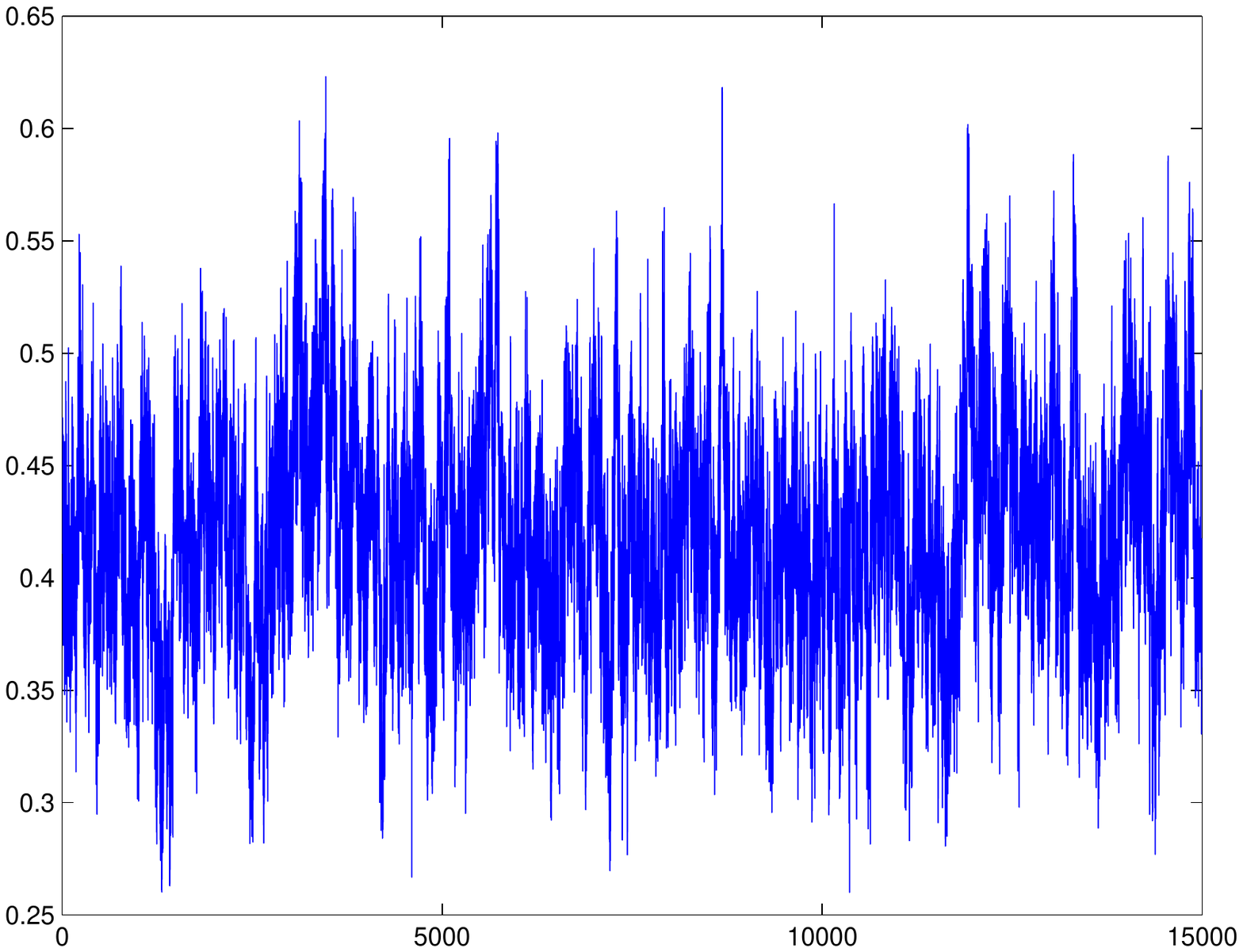}}\quad
\subfigure[Estimated autocorrelations]{\includegraphics[width=5cm,trim=1cm 5cm 1cm 5cm]{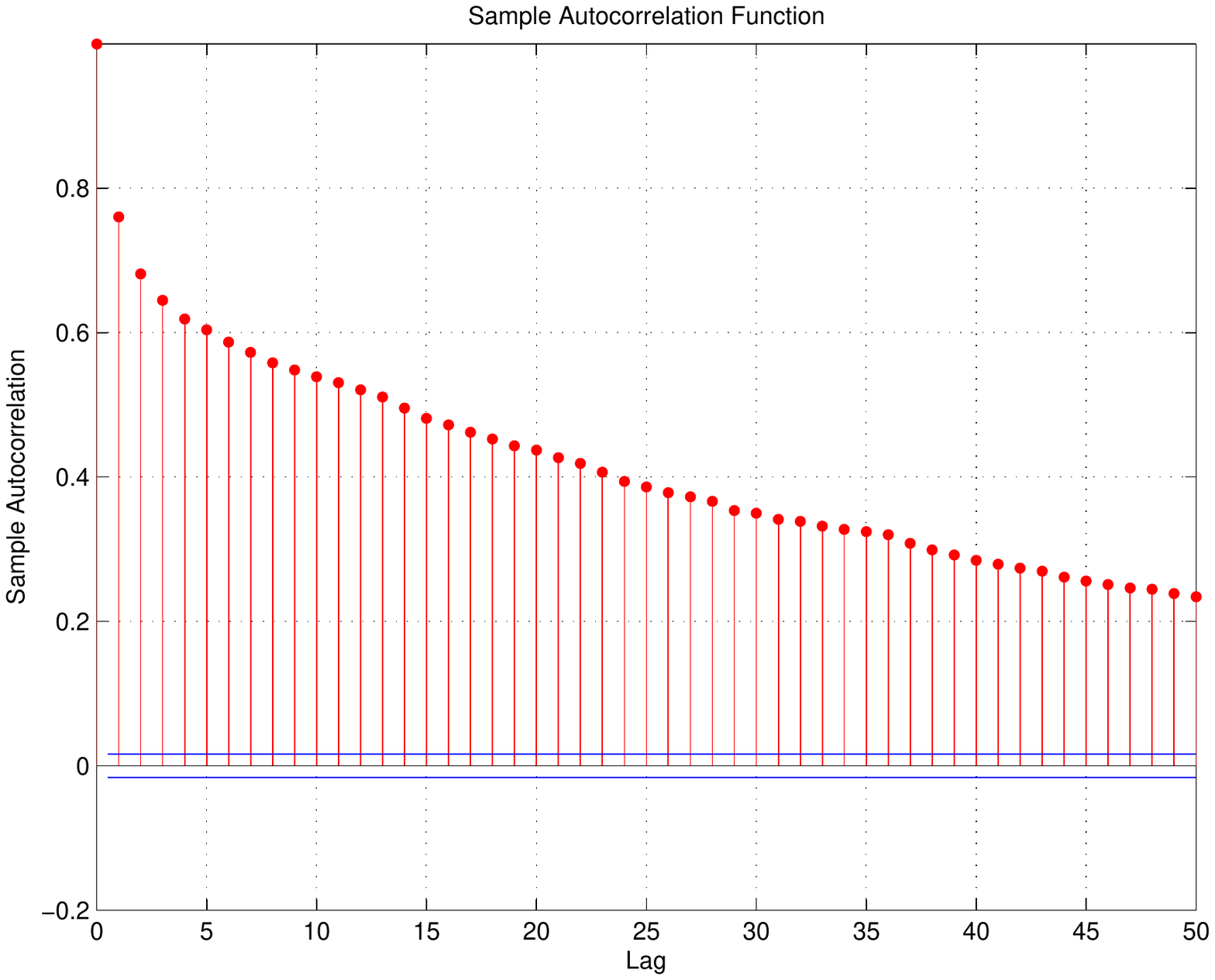} }}
\caption{Analysis of draws of $\beta_{8,2}$ from the estimated GARCH-M model with four latent factors} \label{fig:us17 b}
\end{figure}

The Gibbs sampler produced these results efficiently. The draws of $f$ show a particularly rapid degree of mixing; Figure~\ref{fig:us17 f} shows an example trace plot. The low autocorrelation of the draws is echoed in their low IACT, estimated to be 1.7. The median IACT for all components of $f$ was 2.8, and the maximum was 26. The draws of $\beta$ are a little slower mixing than in the simulated examples considered above, with a typical example plotted in Figure~\ref{fig:us17 b}. The draws of the factor loadings still mix relatively rapidly, with a median IACT of 16 and a maximum of 37.


\section{Conclusion}

Recent developments in sequential Monte Carlo methods have opened up new possibilities for Bayesian computation on latent factor models with time-varying volatility. As our  article demonstrates, the particle Gibbs algorithm provides a flexible and efficient framework for carrying out inference on latent factor models with GARCH factors and GARCH errors. It can be applied to models using an invariant specification for the factor loadings (where possible), or to those using a more traditional triangular identification scheme. The conditionally linear-Gaussian structure of GARCH makes it particularly well suited to particle methods. The resulting parameter estimates mix well and explore the posterior distribution rapidly. It is possible to extend the methodology in a straightforward way to GARCH factor models with regime changes and structural breaks. 

\section{Acknowledgement}
Jamie Hall was  partially supported by ARC grants DP120104014 and LP0774950. Robert Kohn was partially supported by
ARC grant DP120104014.
\bibliographystyle{asa}
\bibliography{FactorGarch}

\end{document}